# HETEROGENEOUS NUCLEATION-CURRENT TRANSIENTS UNDER CHEMICAL REACTION CONTROLL.

P.C.T.D'Ajello, I. Mozolevski and Z.G.S.Kipervaser

Departamento de Física UFSC/CFM
P. O. Box 476 – CEP 88040-900
E-mail:pcesar@fisica.ufsc.br
Fax:++55 48 231 9688

**ABSTRACT**

Heterogeneous nucleation on catalytic surfaces plunged into a fluid is described through a stochastic model. To generate this non-equilibrium process we assume that the turn on of a electrostatic potential triggers a complex dynamics that includes a free Brownian motion, a reaction kinetic and a stimulated migration before the final adhesion of ions on the surface (electrode). At $t = 0$, when the potential is switched on, the spatial symmetry is broken and a two-stage process is developed. First the ion undergoes a change in its electrochemical character (at some region of the space) and then reacts at some specific points to stick together on the surface. The continuous addition of ions develops a material deposit connected to the current transient signals measured in electrochemical deposition processes. Unlike current models found in the literature, this procedure avoids the computation of the area covered by the diffusion zones, allowing a formalism skill to describe equally well the absorption of ions by channels on electrified surfaces. The theory is applied to the electrodeposition of nickel on n-silicon and explains the failure of standard three-dimensional nucleation models to reproduce situations where interfacial relations control the dynamics. It is shown that, processes dominated by chemical reactions on the electrode can not be classified as possessing an exclusively instantaneous or progressive character, rather, they mix together these two limiting forms of nuclei activation to defines the current transients in its beginning.

## I - INTRODUCTION

Nucleation and growth of deposits into well-localized regions of the space, like surfaces plunged inside an initially homogeneous medium, is nowadays a very important subject of academic research. In fact there is a great number of theoretical models describing the current transient occurring in electrochemical deposition events[1-7]. In most cases they are originated as improvements to the original work of Sharifker-Hills[1] (SH) that centralize its attention in two points: the ion regular diffusion near the electrode surface and a scheme to compute the time evolution of the effective reaction area on the material deposit surface. As consequence the models differs, basically, in the way the effective reaction area is described after the "overlap" of the three-dimensional diffusional fields, during the nucleus' evolution. This kind of approach, which we call standard approach from now on, is charged by two problems: i) the emphasis to estimate (in a too high level) the geometric form of the deposits, without an adequate mode to address it and, ii) the persistence to handle of exclusively diffusion dominated electrodeposition processes. In effect, in these models it is generally accepted that the nucleus growth following the diffusional fields' behavior that starts under a hemispherical diffusion control and evolves up to a linear diffusion control (after the complete "overlap" of the individual nucleus). Notwithstanding, these assumption concerns to a model (an intellectual, immaterial conception) and they are not in correspondence with reality. In fact, many experiments has revealed material depositions that show some roughness that grows up until some ten or hundred Angstroms, from bottom to top, when semi-conductor electrodes are used[8]. In these cases the time evolution do not show a surface softness. In a similar way, it is not possible to conceive that all electrochemical deposition process is, exclusively, controlled by diffusion. A diffusion-controlled process implies in a deposition that develops at reaction rates (on the deposit's surface), infinitely high that those related to the concentration changes. It is well known that, not only the reaction rates on the catalytic surfaces can be modified in order to get a slow dynamic (and becoming the dominant process) but also that, the matter transport (diffusion) could be changed by enhancement of electrical and/or chemical effects. It is not difficult to conceive that these enhancements can originates convective movement and also selective reactions that include the discerned ions. The standard approach has given excellent results but the obsession to work with just one model could induce to severe prejudice. In this case it postpone some questions that claim for consideration every time electrodeposition is described, as example: a) the inappropriate usage of the double normalization procedure (current, time) to consider potentiostatic current transients; b) the search for a necessary connection between the potentiostatic current transients and the cyclic voltamogram identified to the same system; c) the way to consider the effect of a chemical potential modification following the inclusion of some composites in solution; how change the current transients and the deposit's morphology?; d) the way to introduce, in the descriptive equations, the surface characteristic (composition and preparation), in order to evaluate the current transient behavior associated to physic and chemical surface's proprieties?; e) it is possible to include a way to estimate the reactive power of the catalytic surface following some component's inclusion?; etc.

In this work we start this program through the derivation of analytic expressions describing the current transient curves measured in electrochemical deposition processes on semiconductor electrodes. We focus attention in a regime controlled by chemical reactions, instead of a diffusion controlled one, developing a effort to avoid the computation of the effective reacting area and the geometry of the deposits in its early stage. We use a two-step model recently introduced[9, 10] and schematically indicated in Figure 1 that sketches the near electrode region, inside an electrolytic cell. We realize that, at distant positions (greater than some cents of a micron) from the electrode, the hydrated ion is assumed to perform a Brownian motion. Closer to the electrode (on the first reaction hemisphere indicated in Figure 1), the Brownian particle feels the effect of a field that changes its electrochemical character. The important point in this transformation is to account for a change in the ions dynamic. The details and the precise nature of this modification are not relevant to the model. After this reaction, the transformed ion, turns to be very sensitive to the field generated near the electrified plane. Then (inside region II in Fig. 1), the ion migrates to the plane where it is reduced before its adsorption on localized points (the nuclei). As consequence of the continuous adhesion of ions, a material deposit grows in close connection to the flux of charges used to reduce the ions.

Following the schema devised in Figure 1, we have to work with a solution obeying different conditions in space. In this sense we adopt the two step course: (1) we obtain the flux of ions, $j_h(R,t)$, through the first reaction hemisphere and, (2) we derive the probability, $p_d(r',t)$, that an ion, which is inside the region II, arrives at a nucleus where it is reduced. The knowledge of these two factors gives us the current density on the electrode or, what is equivalent, the growth rate of the deposit, which reads:

$$I = p_d(r',t).\, j_h(R,t)\,, \qquad (1)$$

where $I$ represents the measured current density.

This model has received a careful introduction in two previous papers[9,10] and it is suitable to discuss the two-limit situation found in electrochemical processes; the diffusion controlled deposition and the reaction controlled deposition[11]. In the first, of this papers[9] , we have assumed an infinitely high rate for ions' transformation on the first reaction hemisphere, such the overall dynamics is controlled by diffusion. In this case the results obtained by Sharifker-Hills (SH)[1] becomes a particular case of the derived current density. A predictable result, once the SH model represents a limit case for a diffusion-controlled processes. It is also verified that, under diffusion control, the progressive and instantaneous nucleation regimes are clearly identified and related to the Avrami[1,2] factor through the probability $p_d(r',t)$ indicated in Equation (1) and given by:

$$p_d(r',t) = (constant).(Avrami \quad growth \quad rate). \qquad (2)$$

In effect, once the interfacial reaction rates are greater than the rate of matter transport, the slower process (the diffusion) regulates the electronic flow. In this case the nuclei activation mechanism has a minor significance and obey a single classification (instantaneous or progressive) to define the role of the interfacial reactions on the entire process. In Fig. 2 we show a graphic representation of the current transient curves in a diffusion-controlled deposition. In the figure caption, the left-hand bracket is connected to $j_h(R,t)$ and the right hands one to the probability $p_d(r',t)$. In this figure it is particularly remarkable the perfect mach between theoretical (including SH curve) and experimental results for $t \le t_{max}$. In this case the factor $t'^2$ in the exponential argument (see equation in the figure caption), characterizes the progressive behavior for a diffusion-controlled nucleation. Similar results are obtained for instantaneous nucleation which, by its turn, is characterized by an exponential argument having a linear time dependence (see reference 9). Also in this case the SH results coincide with the theoretical and experimental results in the range $t \leq t_{max}$, although its poor adequacy for $t \rangle t_{max}$ .

In the second paper[10], we change the boundary conditions on the first reaction hemisphere in order to consider nucleation controlled by interfacial relations. The same schema depicted by Equations (1) and (2) is used but, the flux of ions $j_h(R,t)$, obtained by stochastic considerations, turns to be hard worked. In this case we shown that the SH curves never mach the experimental ones, attesting a model conceived to describe the current transients under interfacial reactions control. In Fig. 3 we show a set of curves that attends for instantaneous nucleation in reaction controlled processes. Once again, the left-hand bracket, in the equation depicted at the figure caption, is connected to $j_h(R,t)$ and the right hands one to the probability $p_d(r',t)$. The theoretical curves produces a reasonable agreement with the SH results for times $\frac{t}{t_{max}} \leq 1$, but differs from the experimental results whose points are located between the two limiting curves of the SH model as we will show during the analysis of Fig. 5. In spite of an adequate description of the current transients in the range $t \geq t_{max}$, where diffusion is important to define the shape of the curves, the disagreement with experimental results at the beginning of the deposition process, when $t \leq t_{max}$, indicates the importance of the nucleus activation mechanism to describe the current transients. Effectively, under chemical reaction control the current transients are so delicately affected by the nucleus activation mechanism that it hardly could be considered as just progressive or instantaneous. The verification of this point of view is the intention of the present work. As in the two previous articles[9,10] we use the two-stage model drafted in Fig. 1 and explained on what follows. We also use equation (1) to generate the current transient curves, which will be compared to experimental results in section IV.

To conclude this introduction we wish to emphasize the intention to prove that, in reaction controlled depositions, the nuclei activation is not exclusively guided by an instantaneous or progressive behavior but rather by a chemical kinetic that mix together the two kind of nuclei evolution in the transient regime. To demonstrate this, we derive a new expression for the $p_d(r',t)$ entering into Eq. (1), such Eq. (2) is no longer valid when deposition and growth obeys an interfacial controlled kinetic.

**II – THE THEORETICAL MODEL**

We work with a medium where a great number of particles are diluted into an electrolytic solution, mainly constituted by water molecules. The assembly of particles forms a set of many species, having different mass, different constitution and different ionic states. Among the myriad of species, we focus attention to one particular kind of particle devised as an hydrated/solvated ion, homogeneously distributed into the bulk of the system. To simplify we call this singular ion a Brownian particle which, encapsulated by a shell of water molecules due to polar attraction, moves inside the electrolytic medium performing a Brownian motion. This complex homogeneous medium undergoes a symmetry break at time, $t = 0$, when an electric potential difference is setting up and the electrode is activated. On the electrode, there are many electroactive points (we call them, nucleus, from now on) which are able to interchange charges with the specific ions we are looking for. Some of these points are automatically activate at $t = 0$; others turn to be activated during the course of time, depending on the physical and chemical conditions imposed on the system.

In the vicinity of the electrode our system is very complex, clearly out of equilibrium and, possibly, affected by many different forces as: electrical, gravitational, buoyancy and generalized thermodynamic forces. The competition of these many effects generates the overall dynamics followed by the ions that we describe through a schema based on the four steps listed below.

1. The Brownian particle (hydrated ion) migrates inside the bulk (region I in Fig. 1) describing a free Brownian motion.
2. Brownian particles, which randomly cross a hemispherical surface, defined by a radius $R$, will suffer a change in its electrochemical character, a distortion or instability of the solvatation shells. As shown in Figure 1 this surface is named first reaction hemisphere and the radius $R$, reaction radius (a critical distance around a nucleus). The rate at which the reaction, that changes the electrochemical ionic character, occurs on the first reaction hemisphere is sensitive to the chemical-physics situation (electrical conductivity of the medium, electrostatic potential on the electrode, pH, composition, etc) but once the process is started the radius $R$ stays constant during the growth process.
3. Inside the first reaction hemisphere (region II in Figure 1) the ion migrates towards an electro-active site performing a diffusive motion assisted by some kind of thermodynamic force.

4. At the elecro-active site, the ions are reduced and there, they stick together. Crossing the first reaction hemisphere is not a sufficient condition for the occurrence of a reduction and consequent aggregation of the reduced ion on the growing nucleus.

Once we have defined the model, some preliminary considerations are now necessary.

(a) It will be assumed a random distribution of nucleus on the electrode

(b) From a theoretical point of view, $R$ is introduced to mimic a cut off of some potential or generalized thermodynamic potential. That is, $R$ is considered as a mean distance, from a nucleus, defining a region where the effect of some field, acting on the reacting species, can no longer be ignored. By assuming this parameter, it is possible to obtain an analytic expression for a non-equilibrium process, taking into account spatial inhomogeneities. In this sense the reaction radius works like a mathematical apparatus allowing for analytic solutions for a complex system. $R$ is not a universal constant and its magnitude changes according the physical situation (in a way not identified yet) without affecting the model.

(c) To solve the problem in a realistic way, it is necessary to choose a physical variable to provide opportunity for a comparison with real situations. Given that the Brownian particles are ions, it seems natural to use the electric current density flowing through the electrode, during the deposition process, as the measurable physical parameter.

(d) Once the motion of the hydrated ions is Brownian, the analysis has to be carried out in a probabilistic manner. We first determine the probability to find a particle at a distance $R$ from the center of the nucleus. This information turns possible to determine the number of Brownian particles crossing the area delimited by the radius $R$ per unit of time (the flux of transformed ions). Then if we know the probability that an ion arrives at a nucleus after entering region II as a transformed ion, we can determine the current through equation (1).

## III – CURRENT TRANSIENT EVALUATION

Following the schema devised in the previous section, we need the flux of probability, $j_h(R,t)$, of ions through the first hemispherical surface and the probability, $p_d(r',t)$, that an ion, which is inside the region II, arrive at a nucleus to be reduced. The knowledge of these two factors gives us the current density as prescribed by equation (1).

**$j_h(R,t)$ derivation**

As stated before, in region I the hydrated ions perform a free Brownian motion. On the surface of the first hemisphere they undergoes a kind of reaction, changing its electrochemical character at a rate $K$ that possess a finite value. This means we are assuming the reaction on the surface of the first hemisphere without a mandatory character, i.e., there is a non-null possibility that Brownian particles crosses the first hemispherical surface without underwent a reaction.

Because in reference 10 we have worked this part of the problem, here we give a brief explanation about its solution. To obtain the probability densities, $p(r,t)$, that a Brownian particle is on the first reaction hemisphere we use the Langevin formalism[13]. By a standard procedure the associated Fokker-Planck (FP) equation in spherical coordinate is generated, namely,

$$\frac{\partial\left(rp(r,t)\right)}{\partial t} = D\frac{\partial^2}{\partial r^2}\left(rp(r,t)\right). \tag{3}$$

In this equation $D$ is the diffusion coefficient of the Brownian particle and it is related to the second moment of the distribution whereas, the probability density is related to the Brownian particles concentration through the expression,

$$p(r,t) = \frac{c(r,t)}{c_o}. \tag{4}$$

In Eq. (4), $c(r,t)$ is the ion concentration at a distance $r$ from the nucleus and $c_o$ its concentration, under equilibrium conditions, at time $t = 0$.

Equation (4) is solved obeying the boundary conditions:

$$Kp(R,t) = 2\pi R^2 D\frac{\partial p(r,t)}{\partial r}\Big|_{r=R}, \tag{5a}$$

$$p(\infty,t) = 1 \qquad \forall t, \tag{5b}$$

and the initial condition:

$$p(r,0) = 1 \qquad \forall \quad r\rangle R. \tag{5c}$$

Condition (5a) expresses the possibility that the Brownian particles, arriving at the first reaction hemisphere, have a non-null probability to avoid a reaction/transformation. That is, only a fraction of the Brownian particles, which arrives on the hemispherical surface, gets a transformation. Condition (5b) means the diffusion-reaction process affects the equilibrium distribution of ions just at finite distances from the reacting surface. Far from the surface, the distribution of ions remains essentially the same as in the equilibrium state. Finally, the initial condition (5c) states that, the probability to find a Brownian particle at any distance $r$ $(r\rangle R)$, at $t = 0$, equals unity.

Under these conditions, Equation (3) is solved as demonstrated in reference 10. The solution $p(r,t)$ is then used to evaluate the flux of ions on the first reaction hemisphere through the prescription,

$$j_h = 2\pi R^2 D c_o zF \frac{\partial p(r,t)}{\partial r}\Big|_{r=R} = c_o KzF\left[\frac{\gamma}{R} + \left(1-\frac{\gamma}{R}\right).\exp\left(\frac{Dt}{\gamma^2}\right).erfc\left(\frac{\sqrt{Dt}}{\gamma}\right)\right], \qquad (6)$$

where $erfc(x)$ means the error function complement and,

$$\gamma = \frac{1}{\left(K/2\pi DR^2\right)+\left(1/R\right)}, \qquad (7)$$

a parameter that is born during the integration of Eq. (3). In Eqs. (6) $z$ stands for the charges number per ion and $F$ is the Faraday constant, which gives to $j_h$ the dimension of a current density.

**Evaluation of $p_d(r',t)$ and ionic flux for new structures**

Instead to use again a Langevin type formalism to describe the particle's dynamic inside region II (see Fig. 1), we use a naive scheme to simplify the acquisition of $p_d(r',t)$. We still consider that a complex dynamics is going on inside region II but, we realize that all the events can be adequately described by statistical arguments, without a precise description of the dynamic.

Because we are working with a reaction-controlled process, the transport kinetic is so intense that, every time a reduction occurs at one point, the environment is restored. However, once the reaction rate on the first reaction hemisphere is slow and the number of particles inside extraordinarily large, when compared to the number of reduced particles during the transient regime, we assume an almost constant density inside this volume. Thus, notwithstanding the dynamic followed by the ions inside region II, the reaction kinetic on the surface regulate the process and a chemical kinetic equation can be used to describe the reduction probability. Let us consider a volume, defined by the surface, which envelops all first reaction hemispheres generated at $t = 0$ (see Fig. 4). This volume $V$ always contains the sources of charges which, are on the surface. Its magnitude not requires a precise definition but needs to be great enough to contain many nuclei spaced among them, in such a way, new nuclei could be activated on the surface during the time evolution. We observe that, a parcel of the particles contained in this volume is located outside the hemispheres conformed by the reaction radius (they are in region $g$ showed in Fig. 4). These are the particles, which could fly inside the previously formed hemispheres or inside to the new ones, when the reactions start to deplete its number on the surface.

To calculate $p_d(r',t)$, we look to this characteristic volume $V$ from which, $N_B$ ions will be withdraw by the reduction reaction during the transient regime. If there is $N_T$ particles inside the characteristic volume at $t = 0$ then the probability that an ion, inside region II, be reduced at the surface, at a particular time is given by:

$$p_d(r',t) \cong \frac{N_B}{N_T} , \tag{8}$$

where $r'$ represents a position inside region II. $N_B$ however, is a function of time and some care is required to derive Eq. (8). Actually we are calling $N_B$ the number of reduced particles but there is many other kind of particles composing the electrolytic solution. Among these the $A$ like particles represents the specie that react on the electrode to produce $B$ like particles. Thus, inside $V$, the total number of particles is given by the sum:

$$N_T = N_A + N_B + N_C + N_D + \ldots.. \tag{9}$$

where, $N_A$ is the number of $A$ particles, $N_B$ the number of $B$ particles, $N_C$ the number of $C$ particles, etc. To work with concentrations we divide (9) by $V$,

$$\frac{N_T}{V} = a + b + \left( \frac{N_C}{V} + \frac{N_D}{V} + \ldots \right), \tag{10}$$

such, $a$ and $b$ are the number of particles of type $A$ and $B$ per unit volume respectively.

In our model every time an $A$ particle (the transformed ion) arrive at a nucleus it could react, receiving electrons from the substrate and transmuting to a $B$ like particles. If we assume the active nucleus as the source of the electrons, the catalytic reaction could be represented by,

$$n + a + ze^- \xrightarrow{k} b + n, \tag{11}$$

where $n$ is the number of nucleus per unit volume of the space and $ze^-$ the electronic charge transferred to a particle $A$ in order to form a $B$ particle. In fact when the $A$ particles react on the surface to produces a $B$ particle, the first is subtracted from the volume whereas the other is added to the substrate. In this sense there is a conservation of the particle number $N_T$. However, to fix the concentration of $A$ particles on the surface, we realize that the high diffusion rate homogenize the medium in very short time intervals. When the $A$ particles react its number is depleted inside one of the many hemispherical surfaces defining region II and another particles are supposed to come in from other regions belonging to the volume $V$ (those indicated by $g$ in Fig. 4) or come in from out there. To maintain constant the number $N_T$ we assume that incoming particles, which initially are outside the volume $V$, are irrelevant during the transient regime. Then, assuming Eq. (11) as an adequate description to the rate of material deposits increment, regulated by the reaction rate, $k$, we could write:

$$\frac{db}{dt} = kna, \tag{12}$$

where $dt$ means an infinitesimal time interval. Eq. (12) can be rearranged using relation (10), to generate an equation for $b$;

$$\frac{db}{\Omega.\left(1-\frac{b}{\Omega}\right)} = kn.dt\,. \qquad (13)$$

$\Omega$ is the constant term,

$$\Omega = \frac{N_T}{V} - \left(\frac{N_C}{V} + \frac{N_D}{V} + \ldots.\right).$$

At this point we introduce a hypothesis about the nuclei activation process to define thew time evolution of $n$. In the simplest case $n = n_0$ is a constant that gives the density of nucleus at $t = 0$, when the electrostatic potential is switched on to produces an instantaneous activation of all nucleus. A more general situation includes a progressive increase of the nucleus density, which, in its simplest form, is a linear function of time,

$$n(t) = n_0 + \alpha t\,, \qquad (14)$$

where $\alpha$ is the activation rate describing the emergence of new nuclei during the transient process. Inserting Eq. (14) into Eq. (13) and performing a simple integration we obtain:

$$b = \frac{N_B}{V} = \frac{N_T}{V}\left[1-\left(\frac{N_C}{N_T} + \frac{N_D}{N_T} + \ldots\right)\right].\left[1-\exp\left(-\left[n_o kt + \frac{\alpha}{2}kt^2\right]\right)\right]. \qquad (15)$$

Multiplying Eq. (15) by $\frac{V}{N_T}$ we arrive at Eq. (8), i.e.:

$$p_d(r',t) = C_1.\left[1-\exp\left(-\left[n_o kt + \frac{\alpha}{2}kt^2\right]\right)\right]. \qquad (16)$$

Eq. (16) gives the probability that an $A$ ion, which is inside the volume $V$, arrives at the surface to undergoes a reaction. This result, together with Eq. (6), is inserted into Eq. (1) to define the current density;

$$I = N.C_2'.\left[\frac{\gamma}{R} + \left(1 - \frac{\gamma}{R}\right).\exp\left(\frac{Dt}{\gamma^2}\right).erfc\left(\frac{\sqrt{Dt}}{\gamma}\right)\right].\left(1 - \exp\left(-\left[n_o kt + \frac{\alpha}{2}kt^2\right]\right)\right), \qquad (17)$$

completing the derivation.

At this point it is important to emphasize the difference between the reaction rates $K$ and $k$ appearing in Eqs. (6) and (16) respectively. If we look to Eqs. (5.a) and (12) we see that $K$ stands for the rate at which the solvated ions change their electrochemical character on the first reaction hemisphere whereas $k$ is the reaction rate that quantifies the velocity that those ions are reduced on the electrode surface. They both are reaction rates but describing different reactions at different places. They also are introduced in a quite different way. $K$ enters through a boundary condition, (Eq. 5.a), relating the probability that a Brownian particles, coming from the bulk of the electrolytic solution, undergoes a conformational reaction on the first reaction hemisphere. By its turn, $k$ is introduced into the description through a chemical kinetic equation, (equation 12), to quantify the frequency at which the electrons are transferred in order to reduce the ions on the second reaction hemisphere (the hemisphere of deposits).

Another point to clarify is about the way the nucleus works after a well-developed deposit has grown up over them. In fact a nucleus is always a source of charges then, when a material deposit lay over it, the charges need to be transferred to a point on the surface of the second reaction hemisphere. Because in some place of this surface are located the ions to be reduced at any time. The charge transfer depends of the deposit electric physical properties. If it is a metallic one, all points on its surface are equivalent and, the charges could be transferred to anyone of them according the location of the reacting particles at the moment the reducing reaction takes place. The point on the surface, to which the charge is transferred, defines the shortest distance between the surface and the ion that react at that moment. There is no distinction among a reaction on the nucleus itself or at any point on the surface of a metallic deposit, because the time spent to transfer the charges from the nucleus to these points is irrelevant compared to the characteristic times of reactions. To represent this idea, Fig. (4) shows a sketch of some deposits in an electrochemical cell. It is realized that only one electroactive point per nucleus exist on the corresponding surface during every reaction. The points marked “S” in Fig (4) represents the locus where a reduction is taking place at a given instant of time.

**IV – RESULTS**

Following the common usage the comparison between the theoretical results and the experimental data are performed through the double normalized current transients. As pointed out at the introduction this procedure gives rise to misunderstanding but we postpone this discussion to a forth-coming article. Then the current is normalized by its maximum value, $I(t_{\max}) = I_{\max}$ whereas the time is normalized by $t_{\max}$ . After this procedure Eq.(17) reads:

$$\left(\frac{I}{I_{max}}\right)^2 = C'\left[\frac{\gamma}{R} + \left(1 - \frac{\gamma}{R}\right)\exp\left(\frac{D't'}{\gamma^2}\right)erfc\left(\frac{\sqrt{D't'}}{\gamma}\right)\right]^2 \left(1 - \exp\left(-\left[n_o k t_{max} t' + \frac{\alpha}{2} k t_{max}^2 t'^2\right]\right)\right)^2 \quad (18)$$

where $C'$ is a constant obtained from Eq. (17) when $t = t_{max}$, and $t'$ is the dimensionless variable $t' = t/t_{\max}$ whose introduction requires, $D' = Dt_{\max}$ .

The parameters $n_o$, $D$, $R$, $K$, $k$ and $\alpha$, must depends on the applied potential, the electrolytic solution and the electrode characteristics. However, up to now an explicit relation between the electric potential difference and some of these variables does not exist; therefore their values are fixed following an examination of the experimental data or through a plausible argumentation. To trace curves (a), (b) and (c) in Fig. 5, we fix the value of the diffusion coefficient adopting a magnitude currently found into the literature[14,15]. The same is true for $n_o$, whose value is in agreement with the number of nuclei per unit area currently assigned to silicon electrodes[16,17]. $R$ is a new parameter whose magnitude and meaning is not clear still now, as discussed in references [1] and [2], so its value is taken in an arbitrary way. Interfacial controlled process requires a reaction kinetic slower than the diffusion kinetic, so, the magnitude of $K$ and $k$ are constrained by the magnitude of the diffusion constant. Then, fixed $R$ and $D$, we examine the ratio $\frac{K}{2\pi DR}$ to define $K$. This ratio compares the frequency of the reactive process and a frequency connected to diffusion (mass transport). As a criterion we forbid this ratio to assume values much higher than one. Although the dominance of interfacial control over diffusion control results from the combined effect of $K$and $k$, when: $\frac{K}{2\pi DR}\langle\ 1$, the process is reaction-controlled (see curve (a) in Fig. 5). When this ratio is grater than one (like in curves (b) and (c) of Fig. 5), there is a tendency of the curves to get closer the SH results. That is a clear indication of and increase in diffusion contribution to defines the overall process. $k$ and $\alpha$ are selected in orders that both factors be comparable in magnitude. This choice gives to instantaneous and progressive activation equal importance to define the argument of the exponential function in Eq. (16). The parameter, $\alpha$ is set free for adjustment with experimental data but, the reactions rate $K$ and $k$ have values (see Table I) compatible with those found in chemical undergraduate textbooks[18,19]. It is important to note that, rather than a good fit, we search for a behavior (on the current curves) which turns evident the existence of a mixture of instantaneous ($n_o \neq 0$) and progressive ($\alpha \neq 0$) nucleation in reaction-controlled depositions.

In Fig. 5 we show three curves obtained from Eq. (18). They are plotted together with three experimental curves (symbols) and the two limit curves given by the SH model. The situation is the same as that considered in reference [10] and depicted at Fig. 3, however there is a basic difference; now we are using expression (16) to represent the reactions on the surface whereas in reference [10], and in Fig.3, we used the Eq. (2) in its place. A consequence of this change is observed when analyzing the current density slopes between $\frac{t}{t_{max}} = 0$ and $\frac{t}{t_{max}} = 1$. The curves are now (in Fig. 5) in good agreement with the experimental data and, are located between the SH theoretical predictions. When $\alpha = 0$ (Fig. 3) there is a feeble adjustment of the theoretical curves with the experimental ones, at times $t \leq t_{max}$, whereas in Fig. 5 (when $\alpha \neq 0$ and $n_o \neq 0$) we see a good adjustment to all $t$. We also observe that the SH curves are not adequate to describe the current transients in this case, a rather foreseen result, given that SH model is developed for a diffusion controlled process.

We rationalize that, reaction-control depositions requires a more subtle characterization in regard to reaction mechanism. The reactions are so slow, compared to mass transport, that a minimal variation in the chemical kinetic implies considerable changes on the current transients. In effect the solution of the kinetic equation (Eq. (12)) is very sensitive to the form of the nuclei activation. According physical/chemical conditions both regimes, instantaneous and progressive, turns to be equally important. On the other hand, when mass transport regulates the dynamic, the chemical interfacial relation is not so relevant. Then the nuclei activation mechanism assumes a minor importance and the nucleation is adequately described as instantaneous or progressive according the dominance of $n_o k$ or $\frac{\alpha}{2} k$ on the exponential argument of Eq. (16).

Finally it is important to remark that Eq. (16) resembles a generalized Avrami' factor[12,20] where a linear term and a quadratic term in $t$ are mixed together into the exponential argument . It is just this association that allows for a suitable fit between theoretical and experimental results in the range $0 \leq \frac{t}{t_{max}} \leq 1$ (see Fig. 5). For $\frac{t}{t_{max}} \geq 1$ the current transient behavior is strongly affected by the $j_h(R,t)$ (Eq. (6)) contribution, which magnify the role played by the diffusion in the overall process.

**ACKNOWLEGMENTS**

The authors are gratefully indebted to the Brazilian agency CAPES by its financial support.

**FIGURE CAPTIONS**

**Figure 1:** A schematic representation of the different regions and surfaces near the electrode, which are relevant for the stochastic model (see text)..

**Figure 2:** Theoretical curves are obtained from the expression $\left(I/I_{max}\right)^2 = E.\left[1+ R/(\pi D't')^{1/2}\right].\left[1-\exp\left(-2\pi A'N_o D't'^2\right)\right]^2$, for a constant value of the product $AN_o$=50x10$^6$1/cm$^2$s. R assumes the following values, (a) – - –3.16x10$^{-4}$ cm; (b)– – – 3.16x10$^{-3}$ cm; (c)· · ·1.16x10$^{-2}$ cm and (d)– · – · – 3.16x10$^{-2}$ cm. For R =3.16x10$^{-1}$ cm the curve coincides with the continuous line, which represents the SH result. $D' = Dt_{max}$ and $t' = t/t_{max}$ and E is a constant value. Symbol shows experimental curves for a solution containing 104 mM $CoSO_4$, 500 mM $Na_2SO_4$ and 500 mM $H_3BO_3$. ▪ ▪ ▪ -1.02 VxSCE, • • • -1.07 VxSCE, -1.09 VxSCE.

**Figure 3:** Theoretical curves are obtained from the expression $\left(I/I_{max}\right)^2 = C'.\left[\gamma/R + \left(1-\gamma/R\right).\exp\left(D't'/\gamma^2\right).erfc\left(D'^{1/2}t'^{1/2}/\gamma\right)\right]^2.\left[1-\exp\left(4\pi ND't'\right)\right]^2$ for a constant value of D=1.x10$^{-5}$cm$^2$/s; R=3.16x10$^{-4}$cm and N=1.x10$^7$ 1/cm$^2$. *K* assumes the following values: (a) – – – 0.5x10$^{-8}$ s$^{-1}$cm$^3$; (b) - - - - 0.5x10$^{-7}$ s$^{-1}$cm$^3$ and (c)– · –1.0x10$^{-7}$ s$^{-1}$cm$^3$. The continuous line (d) and (e) are the SH results for instantaneous and progressive nucleation respectively. C' is a constant value.

**Figure 4:** The sketch represents a lateral vision of a random distribution of growing deposits as well as an upper vision of the electrode where the deposits grow. At every nucleus, identified by the symbol $n$ in the figure, there is associated an instantaneous charge transmission spot (identified by the symbol $s$) on those nuclei that are reducing an ion at that moment. "g" represents a region of the whole volume $V$ occupied by atoms computed on the definition of $N_T$ but outside the attraction basin defined by $R$. "e" is the envelop surface that delimits $V$ .

**Figure 5:** Theoretical and experimental current transient curves. Circles and squares curves corresponds to current transients obtained as function of the deposition potential for a solution containing 1.0M $NiSO_4$, 0.5M $Na_2SO_4$ and 0.4M $H_3BO_3$. The theoretical curves are obtained for $R = 3.16x10^{-4}\,cm$, $D = 1.0x10^{-5}\,cm^2/s$ and $n_o = 0.25x10^7\,cm^{-2}$. The parameters $K,\ \alpha$ and $k$ are defined in Table 1 with (a) –··– ;(b) ......; (c) – – .Curves (e) and (g) are the SH results for instantaneous and progressive nucleation, respectively.

## TABLE CAPTIONS

Table 1: Parameter values used to generate the theoretical curve (a), (b) and (c) in Figure 5.

| Function | (a) | (b) | (c) |
|---|---|---|---|
| | | | |
| K (cm$^3$s$^{-1}$) | 8.5x10$^{-9}$ | 2.28x10$^{-8}$ | 3.17x10$^{-8}$ |
| k (cm$^3$s$^{-1}$) | 1.17x10$^{-4}$ | 1.01x10$^{-4}$ | 1.43x10$^{-4}$ |
| 2α (s$^{-2}$) | 2.05x10$^{10}$ | 1.46x10$^{10}$ | 2.03x10$^{10}$ |

P.C.T. D’Ajello......TABLE 1

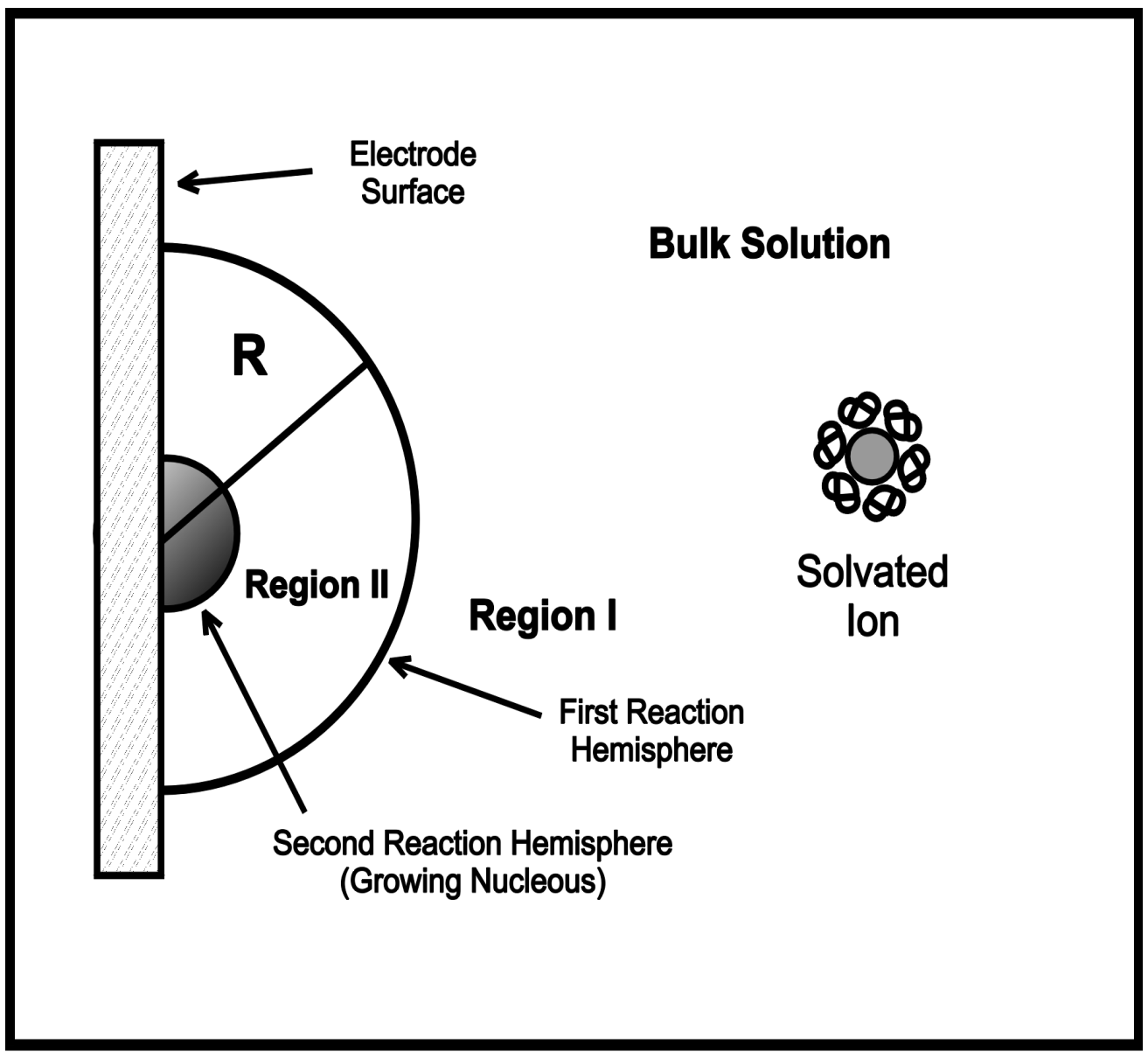

P.C.T.D'Ajello et al.....-Figure 1

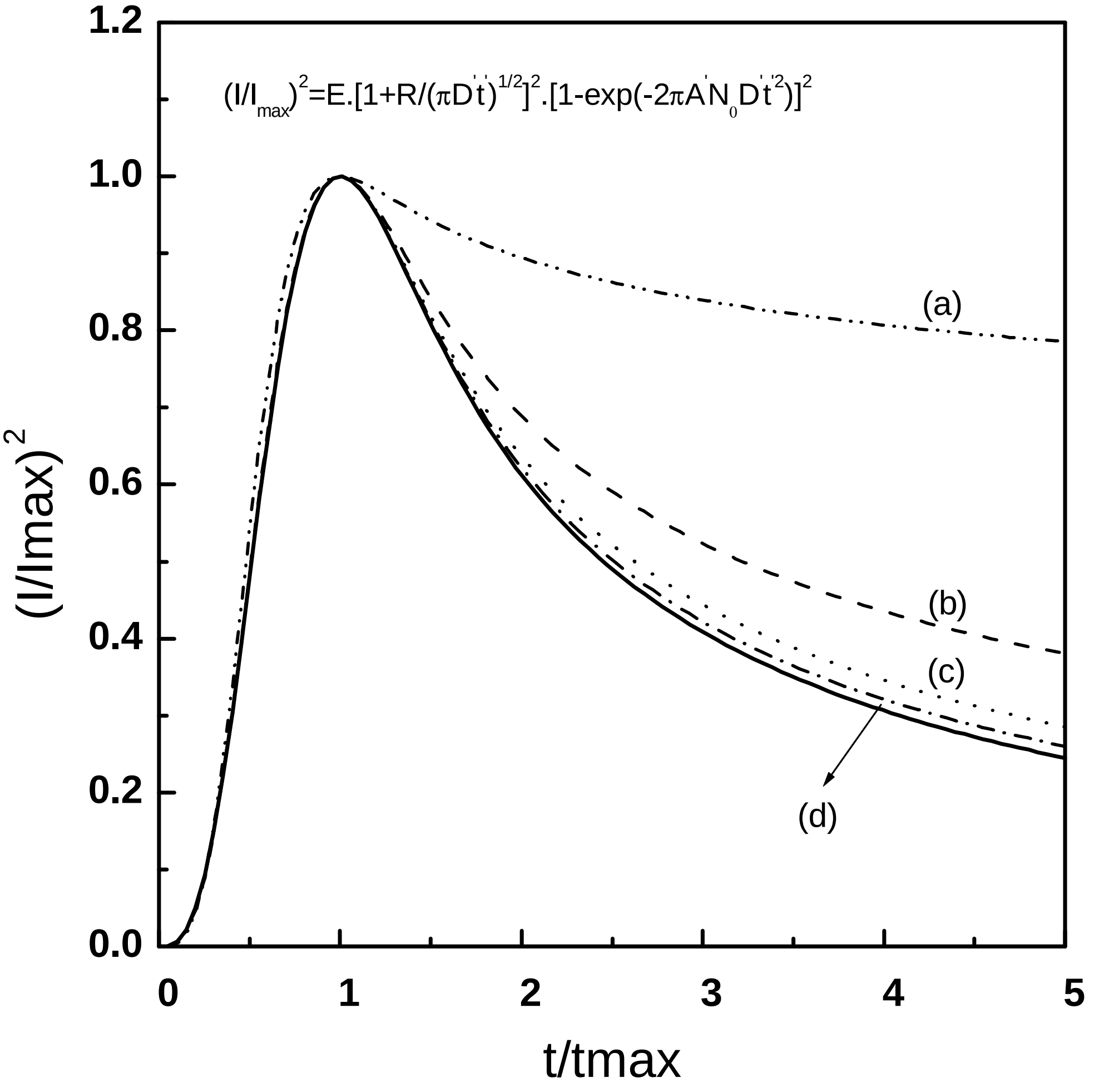

Figure 2 - P.C.T.D'Ajello et al.

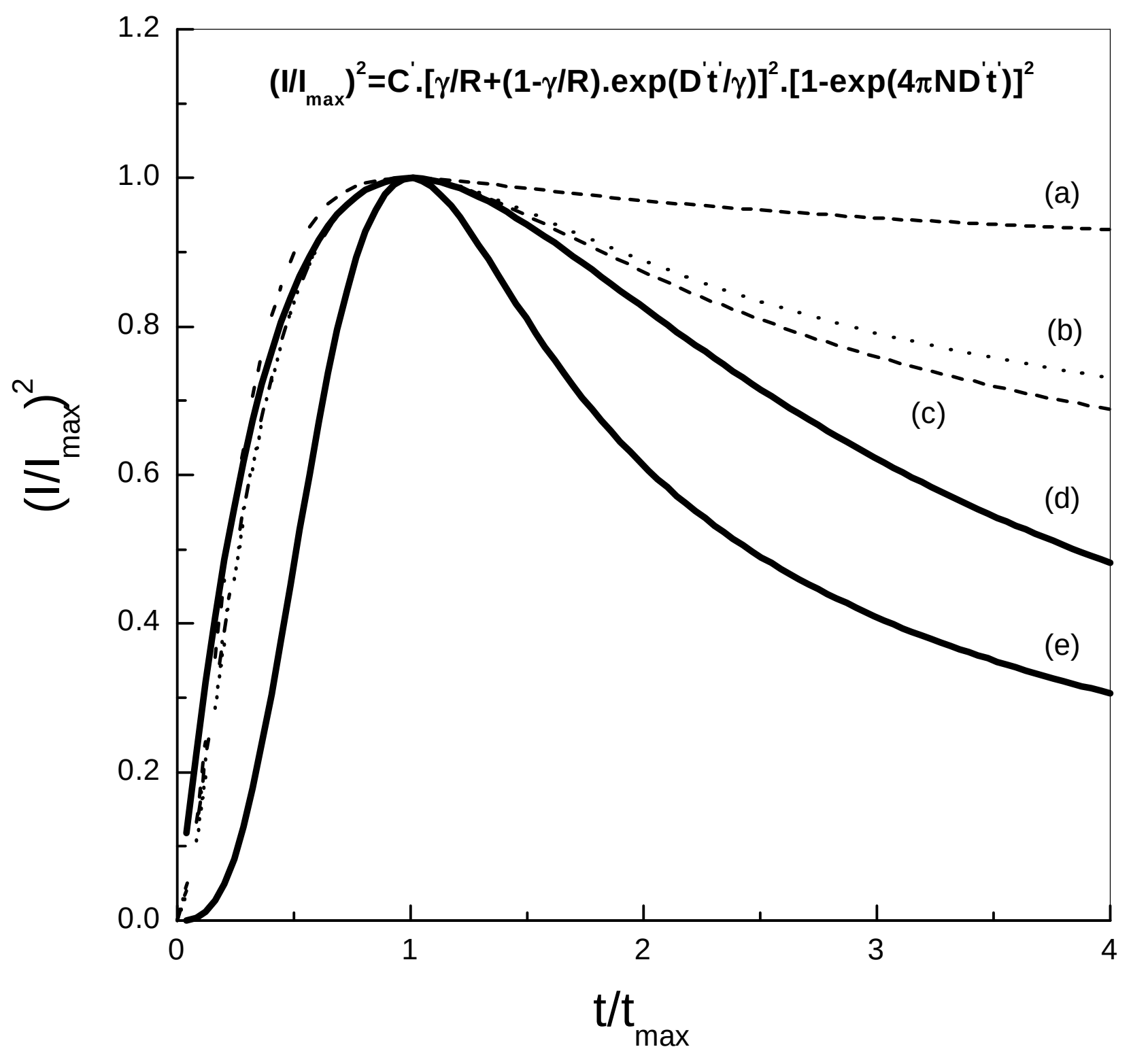

P.C.T.D'Ajello.....Figure 3

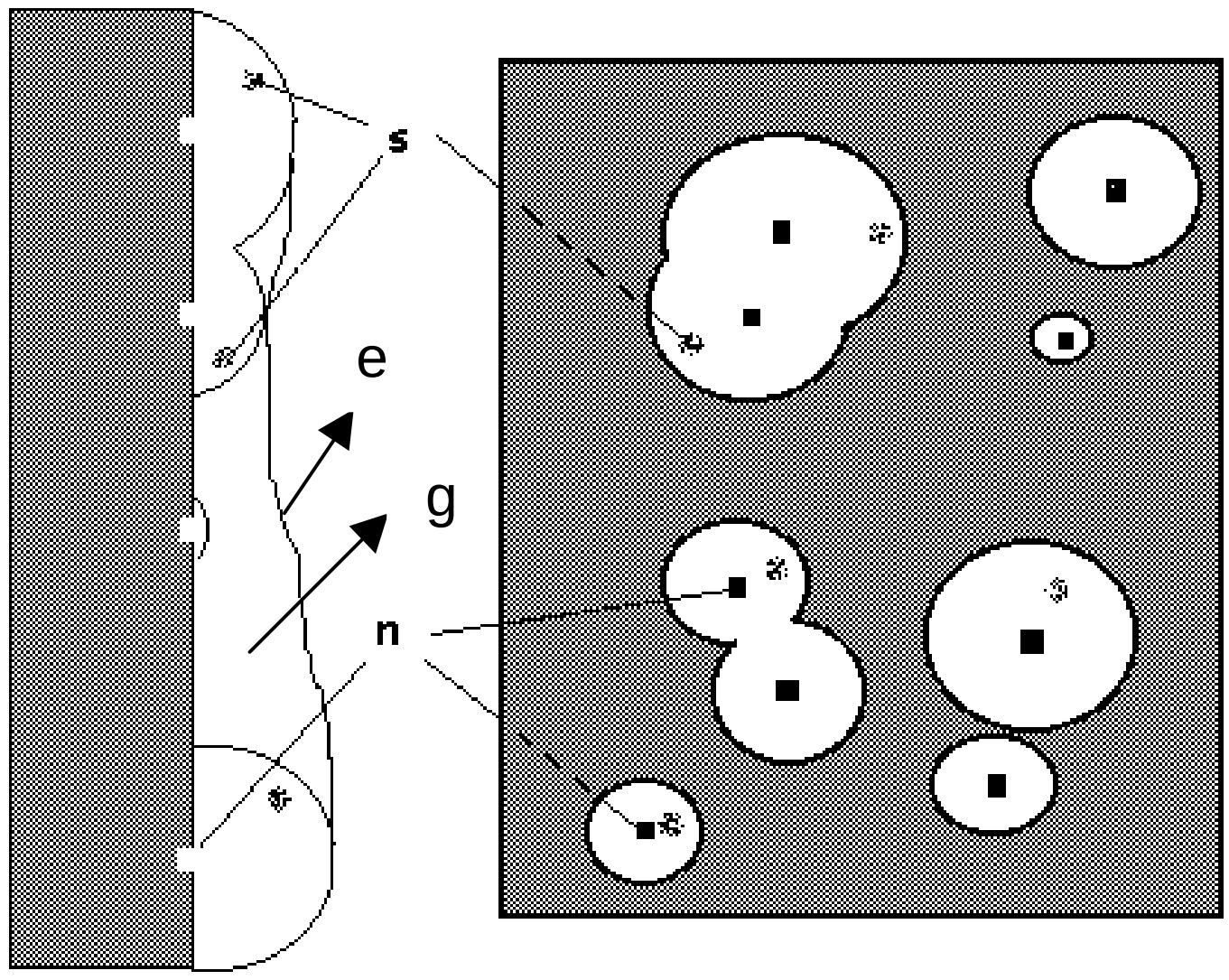

P.C.T. D'Ajello et al.- Figure 4

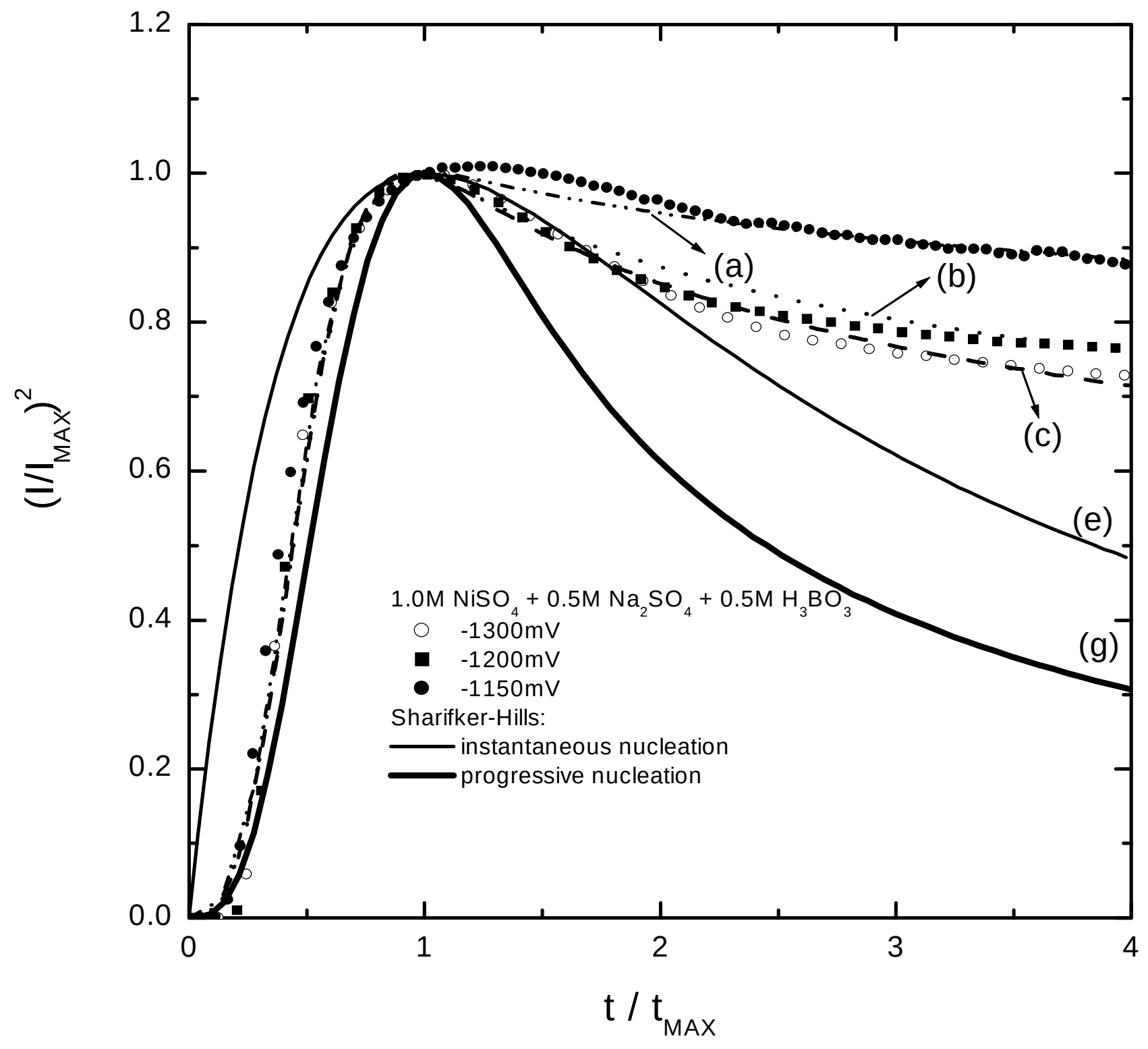

P.C.T.D'Ajello et al.-Figure 5